# Relaying for Multiuser Networks in the Absence of Codebook Information

Ye Tian Aylin Yener
Wireless Communications and Networking Laboratory
Electrical Engineering Department
The Pennsylvania State University, University Park, PA 16802
*yetian@psu.edu* *yener@ee.psu.edu*

June 23, 2018

**Abstract**

This work considers relay assisted transmission for multiuser networks when the relay has no access to the codebooks used by the transmitters. The relay is called oblivious for this reason. Of particular interest is the generalized compress-and-forward (GCF) strategy, where the destinations jointly decode the compression indices and the transmitted messages, and their optimality in this setting. The relay-to-destination links are assumed to be out-of-band with finite capacity. Two models are investigated: the multiple access relay channel (MARC) and the interference relay channel (IFRC). For the MARC with an oblivious relay, a new outerbound is derived and it is shown to be tight by means of achievability of the capacity region using GCF scheme. For the IFRC with an oblivious relay, a new strong interference condition is established, under which the capacity region is found by deriving a new outerbound and showing that it is achievable using GCF scheme. The result is further extended to establish the capacity region of M-user MARC with an oblivious relay, and multicast networks containing $M$ sources and $K$ destinations with an oblivious relay.

This work was presented in part at Asilomar Conference on Signals, Systems, and Computers, Nov, 2011 [1]. This work is supported in part by National Science Foundation under grants 0721445, 0964364 and 0964362.

## I. INTRODUCTION

Relaying is fundamental to the operation in a wireless network. The simplest model that includes this operation is the classical relay channel with one source, one destination and one relay, in the context of which various relaying strategies have been proposed [2], [3]. The capacity of the relay channel is known for special cases, e.g., degraded/reversely degraded [2], and deterministic [3] channels, where the optimality of decode-and-forward (DF) and hash-and-forward is established respectively. For multi-source relay networks, the multiple access relay channel (MARC) and the interference relay channel (IFRC) have been studied in [4]–[14], where various achievable rates and outerbounds have been developed. The capacity for MARC and IFRC are also known under special conditions using DF [4], [12].

A default assumption in all above is that the codebooks used by the sources are known at the relay. In a real system, the codebook can represent the modulation scheme, channel coding scheme and the interleaving pattern used by the sources. In future wireless networks, the wireless devices for different applications may co-exist in the same area sharing the same resources, each having a different codebook, that can change often due to mobility and the time varying nature of the medium. When a relay is used to assist communication between different user pairs, it is required to gather all the codebook information from the sources, in order to perform DF relaying. Exchanging this codebook information can lead to excessive overhead, and inefficient use of wireless bandwidth. It is thus interesting to investigate the fundamental performance limits of a network when the relay(s) does(do) not have access to source codebooks.

To model the uncertainty about the source codebooks at the relay, henceforth called an an oblivious relay, reference [15] has proposed a model which uses randomized encoding at the source, and the codebook becomes common randomness between the source and the destination. The uncertainty about the codebook information at the relay is thus modeled by not informing the relay the common randomness of the codebook. This idea is further investigated in [16], which established the capacity of the *primitive* relay channel with an oblivious relay using compress-and-forward (CF) relaying, where the term *primitive* refers to the relay-destination links being out-of-band and of finite-capacity [3]. The primitive assumption simplifies the model, but retains the ability to characterize the impact of the relay.

For channels with more than one source-destination pairs, for example, the IFRC, when the uncertainty of codebook information is incorporated into the system model, decoding operation can be classified as *interference-aware decoding*, when the interferer's codebook is available at the destinations and *interference-oblivious decoding*, when the interferer's codebook is unavailable at the destinations. Reference [16] has studied the primitive IFRC (PIFRC) with an oblivious relay under interference-oblivious decoding, and established the capacity region using CF relaying and treating interference as noise. For the case of interference-aware decoding, the sum capacity is established using CF relaying, for the special case when the destinations have the same statistics of the received signals [16]. This special case reduces the channel to the primitive MARC (PMARC) with an oblivious relay. However, the capacity region of the PMARC and the general PIFRC under interference-aware decoding remains unknown.

In this paper, we consider this oblivious relaying framework for various multiuser networks and establish capacity results that were previously unknown. The key that renders the results feasible is the achievability scheme. Specifically, we employ a generalized CF (GCF) scheme [17], where the compression index and source messages are decoded jointly instead of sequentially as in the conventional CF scheme. For the PMARC with an oblivious relay, we derive new outerbounds and show that GCF indeed achieves the entire capacity region. For the PIFRC with an oblivious relay under interference-aware decoding at the destinations, we propose a new strong interference condition, under which we derive new outerbounds and show that the capacity region can be established using GCF scheme as well. The capacity results are further extended to the $M$-user PMARC with an oblivious relay and $M$-source $K$-destination multicast network with an oblivious relay.

The remainder of the paper is organized as follows: Section II introduces the system model. Section III establishes the capacity results for MARC with an oblivious relay and multicast networks with an oblivious relay. Section IV establishes the capacity results for IC with an oblivious relay under strong interference conditions. Section V concludes the paper.

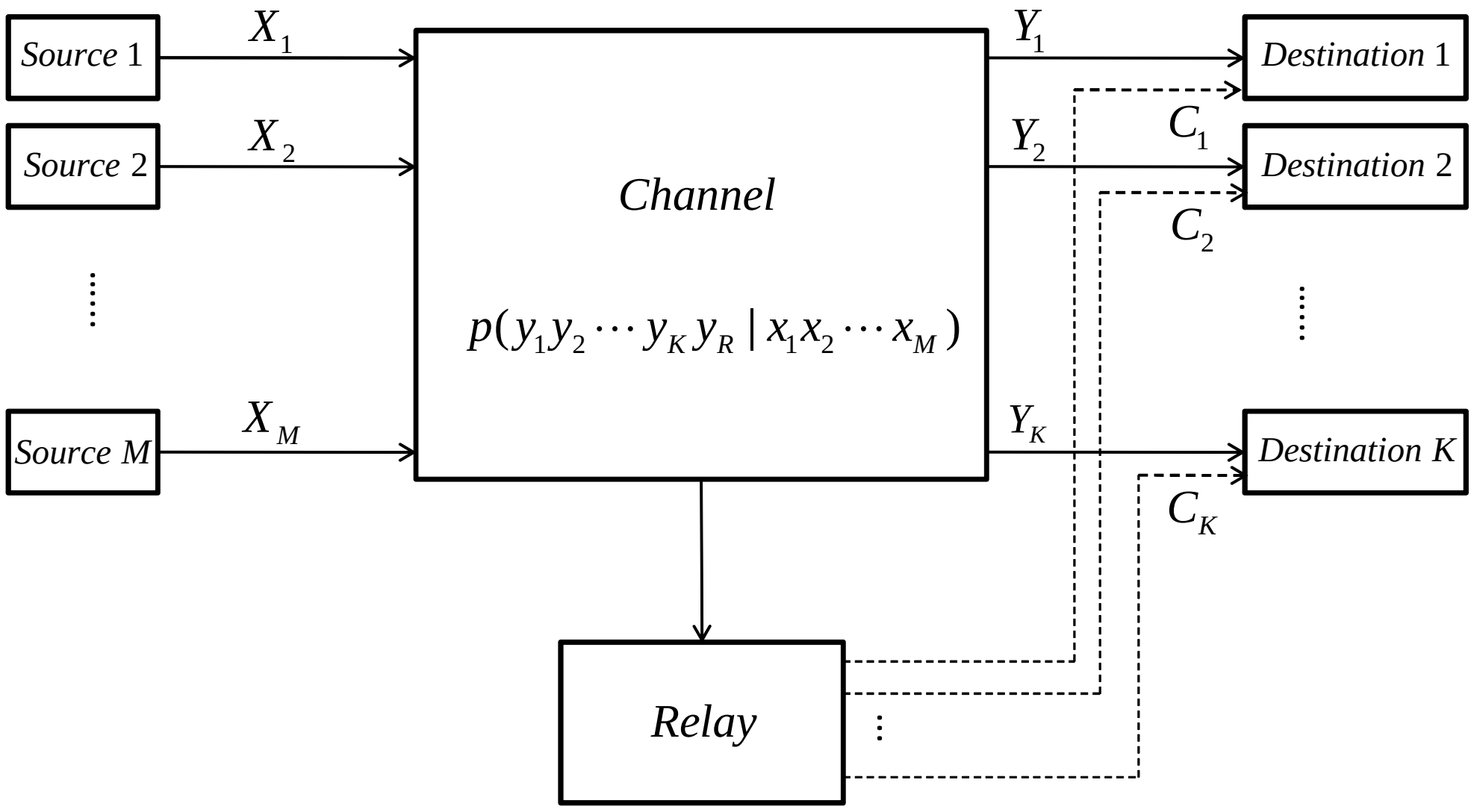

Fig. 1. Primitive multiuser network with an oblivious relay.

## II. System Model

We consider a multiuser network with $M$ sources and $K$ destinations assisted by an oblivious relay, as shown in Figure 1. The relay-to-destination links are digital, i.e., orthogonal with finite capacity $C_i$, $i = 1, \cdots, K$ for the $K$ destinations as shown in Figure 1. We shall refer to this network as a *primitive* multiuser network. The model is then specified into MARC and IFRC in the subsequent sections. We define $\mathcal{M} = \{1, 2, \cdots, M\}$ as the index set of the sources and $\mathcal{K} = \{1, 2, \cdots, K\}$ as the index set of the destinations. We define the message set at source $i$ as $\mathcal{W}_i = \{1, 2, \cdots, 2^{nR_i}\}$, and the alphabet for source $i$ as $\mathcal{X}_i$. We also define the set of codebooks at source $i$ as all possible combinations of length $n$ codewords for each message, where the codewords consist of symbols chosen from the alphabet. The number of codebooks is thus $|\mathcal{X}_i|^{n2^{nR_i}}$ for source $i$. We define an index set of all the codebooks for source $i$ as $\mathcal{F}_i = \{1, 2, \cdots, |\mathcal{X}_i|^{n2^{nR_i}}\}$.

### A. Encoders

We follow the definition in [16] to allow time sharing. We define a $(n, R_1, R_2, \cdots, R_M)$ code for the $M$-source $K$-destination channel assisted by an oblivious relay with time-sharing as

$$(P_{F_i|Q^n}, \phi_i^n) \quad i \in \mathcal{M}, \tag{1}$$

where $P_{F_i|Q^n}(f_i|q^n)$ is the probability of choosing the codebook $f_i \in \mathcal{F}_i$ conditioned on the time sharing sequence $q^n \in \mathcal{Q}^n$. $\phi_i^n$ is the encoding function such that $x_i^n = \phi_i^n(w_i, f_i)$, where $w_i \in \mathcal{W}_i$. The probability of selecting the codebook $f_i$ for source $i$ conditioned on $q^n$ is

$$P_{F_i|Q^n}(f_i|q^n) = \prod_{w_i=1}^{2^{nR_i}} P_{X_i^n|Q^n}(\phi_i^n(w_i, f_i)|q^n) \tag{2}$$

where $P_{X_i^n|Q^n}(x_i^n|q^n) = \prod_{t=1}^{n} P_{X_i|Q}(x_{i,t}|q_t)$.

Note that the codebook and the message are selected independently, i.e.,

$$P_{F_iW_i|Q^n}(f_i, w_i|q^n) = P_{F_i|Q^n}(f_i|q^n) \cdot 2^{-nR_i}. \tag{3}$$

Based on this formulation, we have the following lemma:

*Lemma 1:*

$$P_{X_i^n|Q^n}(x_i^n|q^n) = \prod_{j=1}^{n} P_{X_i|Q}(x_{i,j}|q_j), \tag{4}$$

$$P_{Y_i^n|Q^n}(y_i^n|q^n) = \prod_{j=1}^{n} P_{Y_i|Q}(y_{i,j}|q_j) \tag{5}$$

*Proof:* This relation can be derived following the steps of *Lemma 1* in [15], and the details of the derivation are provided in Appendix A for completeness. ■

The above lemma states that, without the codebook information, the destination sees the transmitted sequence from the source and its received sequence as generated independently from each symbol, given the time sharing sequence. Therefore the received sequence and the transmitted sequence have no structure when no codebook information is available. Note that the randomized selection of the codebook is only to model the uncertainty of the codebook at oblivious nodes. A more detailed elaboration on the codebook information can be found in [15] and *Remark 2-4* in [16].

## B. Channel

The channel is discrete memoryless, and consists of $M$ input alphabets and $K+1$ output alphabets, a channel transition probability, and $K$ out-of-band finite capacity links from the relay to the destinations, i.e.,

$$\mathcal{X}_{\mathcal{M}}, p(y_1 y_2 \cdots y_K y_R | x_1 x_2 \cdots x_M), \mathcal{Y}_{\mathcal{K}}, \mathcal{Y}_R, C_{\mathcal{K}}. \tag{6}$$

## C. Relay Encoder

The relay does not know the codebooks used by the sources. It communicates to each destination with an out-of-band finite capacity link. The messages are generated according to an encoding function

$$\phi_R^n : \mathcal{Y}_R^n \times \mathcal{Q}^n \to \mathcal{S}_1 \times \mathcal{S}_2 \cdots \mathcal{S}_K \tag{7}$$

with $\mathcal{S}_k = \{1, 2, \cdots, 2^{nC_k}\}, k \in \mathcal{K}$. We denote $(S_1, S_2, \cdots, S_K) = \phi_R^n(Y_R^n | q^n)$ as the messages generated by the relay.

## D. Decoders

We assume that the destinations know all the codebooks used by the sources. We consider both *multicast* (MC) and *unicast* (UC) transmissions. For *multicast transmission*, i.e., each source wishes to transmit a message to all destinations, we define the decoding function at destination $j \in \mathcal{K}$ as

$$g_j^{MC} : \mathcal{Q}^n \times \mathcal{F}_1 \times \mathcal{F}_2 \cdots \mathcal{F}_M \times \mathcal{S}_j \times \mathcal{Y}_j^n \to \mathcal{W}_1 \times \mathcal{W}_2 \cdots \mathcal{W}_M. \tag{8}$$

A set of rates $(R_1, R_2, \cdots, R_M)$ is achievable if there exists $(P_{F_i|Q^n}, \phi_i^n)$ for all $i \in \mathcal{K}$ such that $(\hat{W}_1^i, \cdots, \hat{W}_M^i) = g_i^{MC}(Q^n, F_1, \cdots, F_M, S_i, Y_i^n)$, and $\Pr\{\cup_{i=1}^M \cup_{j=1}^K \hat{W}_i^j \neq W_i\} \to 0$ as $n \to \infty$.

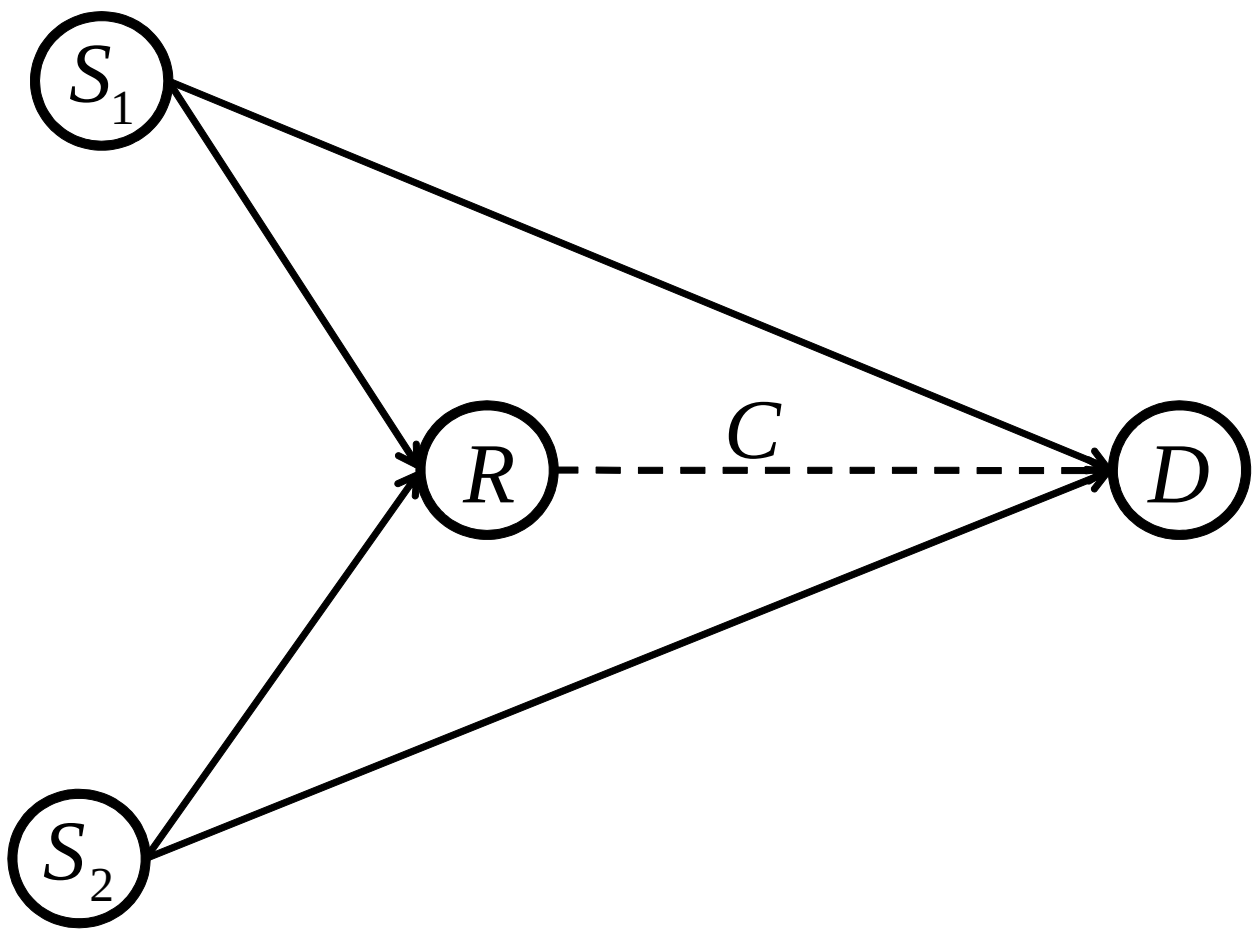

Fig. 2. Primitive multiple access relay channel with an oblivious relay.

For *unicast transmission*, each source only wishes to transmit a message to its intended destination. We assume that $\mathcal{M} = \mathcal{K}$. We define the decoding function at destination $j \in \mathcal{M}$ as

$$g_j^{UC} : \mathcal{Q}^n \times \mathcal{F}_1 \times \mathcal{F}_2 \cdots \mathcal{F}_M \times \mathcal{S}_j \times \mathcal{Y}_j^n \to \mathcal{W}_j. \quad (9)$$

A set of rates $(R_1, R_2, \cdots, R_M)$ is achievable if there exists $(P_{F_i|Q^n}, \phi_i^n)$ for all $i \in \mathcal{K}$ such that $\hat{W}_i = g_i^{UC}(Q^n, F_1, \cdots, F_M, S_i, Y_i^n)$, and $\Pr\{\cup_{i=1}^M \hat{W}_i \neq W_i\} \to 0$ as $n \to \infty$.

## III. Capacity Region for the Primitive Multiple Access Channel with an Oblivious Relay

In this section, we study the PMARC with an oblivious relay. To begin with, we investigate the channel with two sources, one destination, and one oblivious relay, i.e., $M = 2$, $K = 1$ as in the definition of multicast transmission in Section II. The relay-destination link is out-of-band with finite capacity $C$, and the message transmitted from relay to the destination is denoted as $S$. The sum capacity of this channel is obtained in [16], where the CF strategy from [2] with sequential decoding of the compression index and source messages is shown to be *sum-rate optimal*. By deriving new outerbounds, we show that, for the entire capacity region, the GCF scheme is optimal.

### A. Main Result

*Theorem 1:* The following rate region is the capacity region of PMARC with oblivious relaying:

$$R_1 < I(X_1; \hat{Y}_R Y | X_2 Q) \tag{10}$$

$$R_1 < I(X_1; Y | X_2 Q) + C - I(Y_R; \hat{Y}_R | X_1 X_2 Y Q) \tag{11}$$

$$R_2 < I(X_2; \hat{Y}_R Y | X_1 Q) \tag{12}$$

$$R_2 < I(X_2; Y | X_1 Q) + C - I(Y_R; \hat{Y}_R | X_1 X_2 Y Q) \tag{13}$$

$$R_1 + R_2 < I(X_1 X_2; \hat{Y}_R Y | Q) \tag{14}$$

$$R_1 + R_2 < I(X_1 X_2; Y | Q) + C - I(Y_R; \hat{Y}_R | X_1 X_2 Y Q) \tag{15}$$

for all distributions

$$p(q)p(x_1|q)p(x_2|q)p(\hat{y}_R|y_R q)p(y y_R|x_1 x_2) \tag{16}$$

*Proof:* The achievability can be obtained using GCF relaying. We defer the details to Appendix B. For outerbounds, we need to utilize the property (4) and appropriately define the random variable $\hat{Y}_{Ri}$. To illustrate the approach and keep the clarity of the proof, we only present the proof for the individual rates. The proof for the sum rate bound follows from a similar approach and is deferred to Appendix B. For the individual rate $R_1$, we have

$$nR_1 = H(W_1) \tag{17}$$

$$= H(W_1|Q') \tag{18}$$

$$\leq I(W_1; Y^n S F_1 F_2 | Q' W_2) + n\epsilon_n \tag{19}$$

$$= I(W_1; F_1 F_2 | Q' W_2) + I(W_1; Y^n S | Q' W_2 F_2 F_1) + n\epsilon_n \tag{20}$$

$$\leq I(F_1 W_1; Y^n S | Q' W_2 F_2) + n\epsilon_n \tag{21}$$

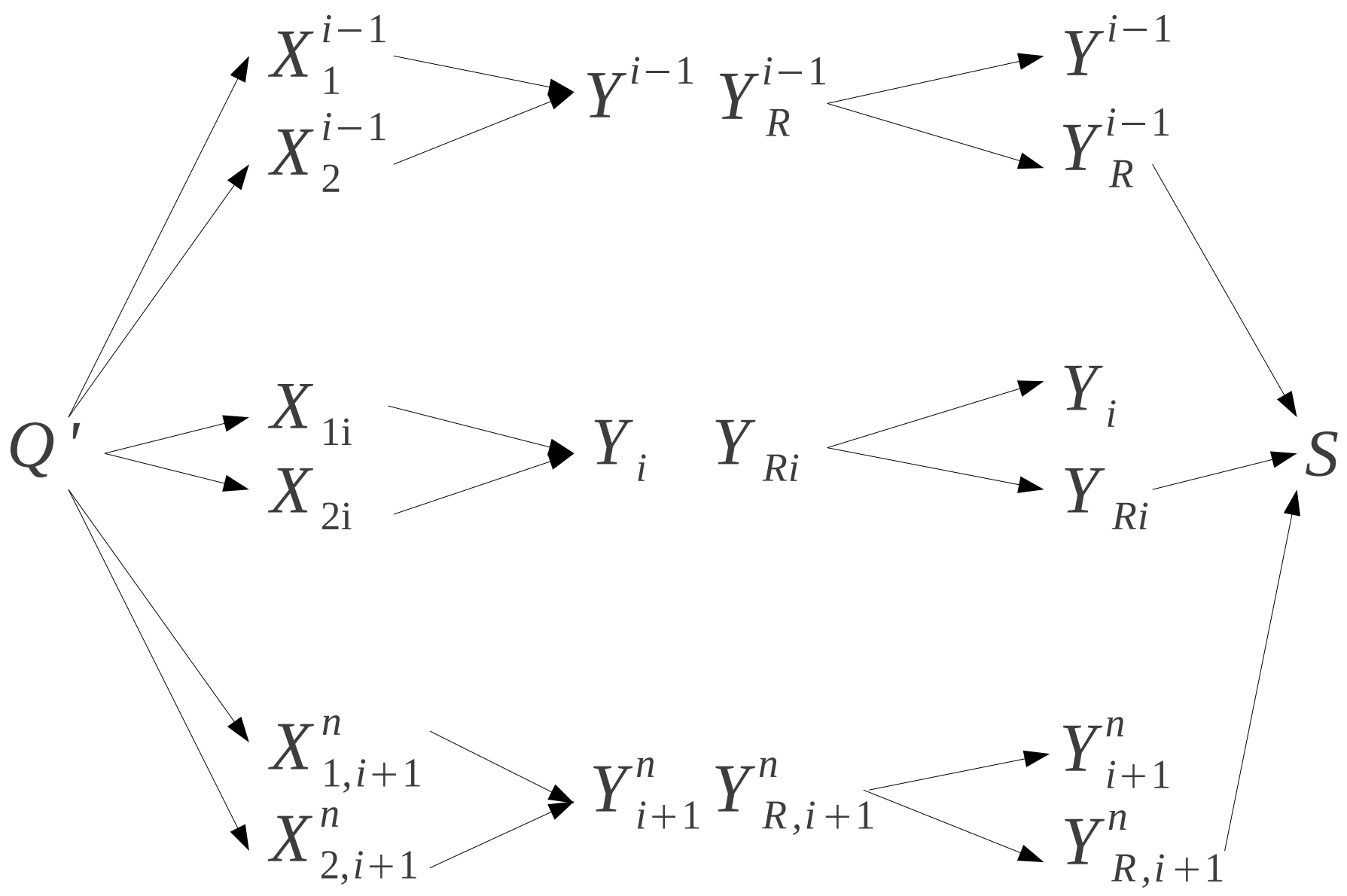

Fig. 3. Markov chain between random variables.

$$\leq I(X_1^n; Y^n S | Q' X_2^n) + n\epsilon_n \quad (22)$$

where $Q' = Q^n$, $\epsilon_n \to 0$ as $n \to \infty$ and (20) follows from the independence between $W_1$ and $F_1 F_2$. Due to *Lemma 1*, we have that the symbols from source sequences are independent for each instance $i$ conditioned on $Q'$ without conditioning on the codebook information. When we combine this property and the memoryless property of the channel, together with the definition of relay encoder, we have the Markov chain between the random variables illustrated in Figure 3, based on the graphical method described in [18, pages 166-168]. Note that we do not consider messages $W_1, W_2$ in the figure since the channel outputs only depend on the channel inputs, and we can upperbound (21) by (22). Without conditioning on the codebooks, we have *Lemma 1* and it is easy to see that the message and source sequence are independent. It thus suffices to consider the random variables shown in Figure 3.

We can further bound (22) with two different methods.

$$I(X_1^n;Y^nS|Q'X_2^n) \tag{23}$$

$$= H(X_1^n|Q'X_2^n) - H(X_1^n|Y^nSQ'X_2^n) \tag{24}$$

$$= \sum_{i=1}^{n} H(X_{1i}|X_{2i}Q') - \sum_{i=1}^{n} H(X_{1i}|Y^nSQ'X_2^nX_1^{i-1}) \tag{25}$$

$$\leq \sum_{i=1}^{n} H(X_{1i}|X_{2i}Q') - \sum_{i=1}^{n} H(X_{1i}|SX_1^{i-1}X_{1,i+1}^nX_2^{i-1}X_{2,i+1}^nY^{i-1}Y_{i+1}^nY_R^{i-1}X_{2i}Y_iQ') \tag{26}$$

$$= \sum_{i=1}^{n} H(X_{1i}|X_{2i}Q') - \sum_{i=1}^{n} H(X_{1i}|\hat{Y}_{Ri}X_{2i}Y_iQ') \tag{27}$$

$$= \sum_{i=1}^{n} I(X_{1i};\hat{Y}_{Ri}Y_i|X_{2i}Q') \tag{28}$$

where (25) follows from *Lemma 1* and the fact that there is no cooperation between the sources. In (27), since conditioning reduces entropy, we introduce some additional random variables in the condition of the second term to form $\hat{Y}_{Ri}$, which is defined as

$$\hat{Y}_{Ri} = SX_1^{i-1}X_{1,i+1}^nX_2^{i-1}X_{2,i+1}^nY^{i-1}Y_{i+1}^nY_R^{i-1}. \tag{29}$$

We can also bound the term (22) in the following way:

$$I(X_1^n;Y^nS|Q'X_2^n) \tag{30}$$

$$= I(X_1^n;Y^n|Q'X_2^n) + I(X_1^n;S|Q'X_2^nY^n) \tag{31}$$

$$= H(Y^n|Q'X_2^n) - H(Y^n|Q'X_2^nX_1^n) + H(S|Q'X_2^nY^n) - H(S|Q'X_1^nX_2^nY^n) \tag{32}$$

$$\leq \sum_{i=1}^{n} H(Y_i|X_{2i}Q') - \sum_{i=1}^{n} H(Y_i|X_{1i}X_{2i}Q') + H(S)$$
$$- (H(S|Q'X_1^nX_2^nY^n) - H(S|Q'X_1^nX_2^nY^nY_R^n)) \tag{33}$$

$$\leq \sum_{i=1}^{n} I(X_{1i};Y_i|X_{2i}Q') + nC - I(S;Y_R^n|Q'X_1^nX_2^nY^n) \tag{34}$$

$$= \sum_{i=1}^{n} I(X_{1i};Y_i|X_{2i}Q') + nC - \sum_{i=1}^{n} I(S;Y_{Ri}|Q'X_1^nX_2^nY^nY_R^{i-1}) \tag{35}$$

$$= \sum_{i=1}^{n} I(X_{1i};Y_i|X_{2i}Q') + nC - \sum_{i=1}^{n}\Big(H(Y_{Ri}|Q'X_1^nX_2^nY^nY_R^{i-1})$$
$$- H(Y_{Ri}|SQ'X_1^nX_2^nY^nY_R^{i-1})\Big) \tag{36}$$
$$= \sum_{i=1}^{n} I(X_{1i};Y_i|X_{2i}Q') + nC - \sum_{i=1}^{n}\Big(H(Y_{Ri}|Q'X_{1i}X_{2i}Y_i)$$
$$- H(Y_{Ri}|SX_1^{i-1}X_{1,i+1}^nX_2^{i-1}X_{2,i+1}^nY^{i-1}Y_{i+1}^nY_R^{i-1}X_{1i}X_{2i}Y_iQ')\Big) \tag{37}$$
$$= \sum_{i=1}^{n} I(X_{1i};Y_i|X_{2i}Q') + nC - \sum_{i=1}^{n} I(\hat{Y}_{Ri};Y_{Ri}|X_{1i}X_{2i}Y_iQ') \tag{38}$$

where (37) follows from the Markov chain in Figure 3, i.e., conditioned on $Q'$, there is no path between the random variables with different time instances. Note that the random variable $S$ and the related edges should not be considered here since $S$ is not present in the third term in (36) and (37) (see [18], pages 166-168 for detailed explanation). The end result can be obtained by introducing another time sharing random variable $Q'' \sim \mathcal{U}(\{1,2,\cdots,n\})$ and setting $Q = (Q'', Q')$. The way we define the random variable $\hat{Y}_{Ri}$ implies the Markov chain $X_{1i}X_{2i} \to Y_{Ri}Q' \to \hat{Y}_{Ri}$ and thus the distribution (16). The individual rate $R_2$ can be obtained in a similar fashion. ■

*Remark 1:* It can be shown that the achievable rate region obtained from GCF includes the one obtained by CF. Specifically, these two schemes have the same maximum sum rate and individual rates, but the rate region due to GCF is potentially larger than that of CF. In fact, using GCF scheme, the random variable $\hat{Y}_R$ can be chosen from a larger set, which leads to a potential improvement in terms of rate region.

*Remark 2:* The noisy network coding scheme proposed in reference [19] is observed to achieve larger rate regions for some multiuser networks. However, for PMARC, noisy network coding has the same achievable rate region as GCF scheme. The reason is that we only have one digital relay-destination link. This limits the gain of noisy network coding, which is provided by relaying without binning, repetition coding and joint decoding. We present the mathematical details related to this remark in Appendix C.

### B. Extension to $M > 2$ Nodes

Using the GCF relaying, and the outerbounding techniques we used in *Theorem 1*, we can derive the capacity region of the M-user MAC, i.e., a channel with $M$ sources and $K = 1$ destination as in the definition of multicast transmission in Section II:

*Corollary 1:* Let $\mathcal{S} \subseteq \mathcal{M}$, $R_{\mathcal{S}} \triangleq \sum_{i \in \mathcal{S}} R_i$. The capacity region of the M-user MAC can be specified by the rate vector $R_{\mathcal{M}}$ satisfying

$$R_{\mathcal{S}} \leq I(X_{\mathcal{S}}; \hat{Y}_R Y | X_{\mathcal{S}^C} Q) \quad (39)$$

$$R_{\mathcal{S}} \leq I(X_{\mathcal{S}}; Y | X_{\mathcal{S}^C} Q) + C - I(\hat{Y}_R; Y_R | X_{\mathcal{M}} Y Q), \quad (40)$$

for $\forall \mathcal{S}$ and all distributions

$$p(q) \prod_{i=1}^{M} p(x_i|q) p(\hat{y}_R | y_R q). \quad (41)$$

The proof is an extension of *Theorem 1*. Note that for the M-user MAC with one relay, the increase of the achievable rate is at most the capacity of the out-of-band link between the relay and the destination, i.e., when the negative term of (40) is 0, and (40) is the dominating term. For this ideal setting, the rate improvement of $C$ bits is shared by all the M-users. Note that as the number of users increases, the average rate improvement for all the $M$ sources, which is $\frac{C}{M}$, becomes negligible. This issue can be resolved by introducing multiple relays using GCF relaying.

## IV. Capacity Region for Multi-Destination Networks with an Oblivious Relay

### A. Capacity Region for the Primitive Interference Relay Channel with an Oblivious Relay

In this section, we present a result for the primitive interference relay channel (PIFRC) with an oblivious relay, where the relay connects to destination 1 (2) with an out-of-band link with capacity $C_1$ ($C_2$) and the destinations are assumed to be interference-aware. The signals sent from the relay to the destinations are denoted as $S_1$ and $S_2$. The channel model is shown in 4. We first establish a new strong interference condition, and then derive outerbounds to show that GCF achieves the capacity region.

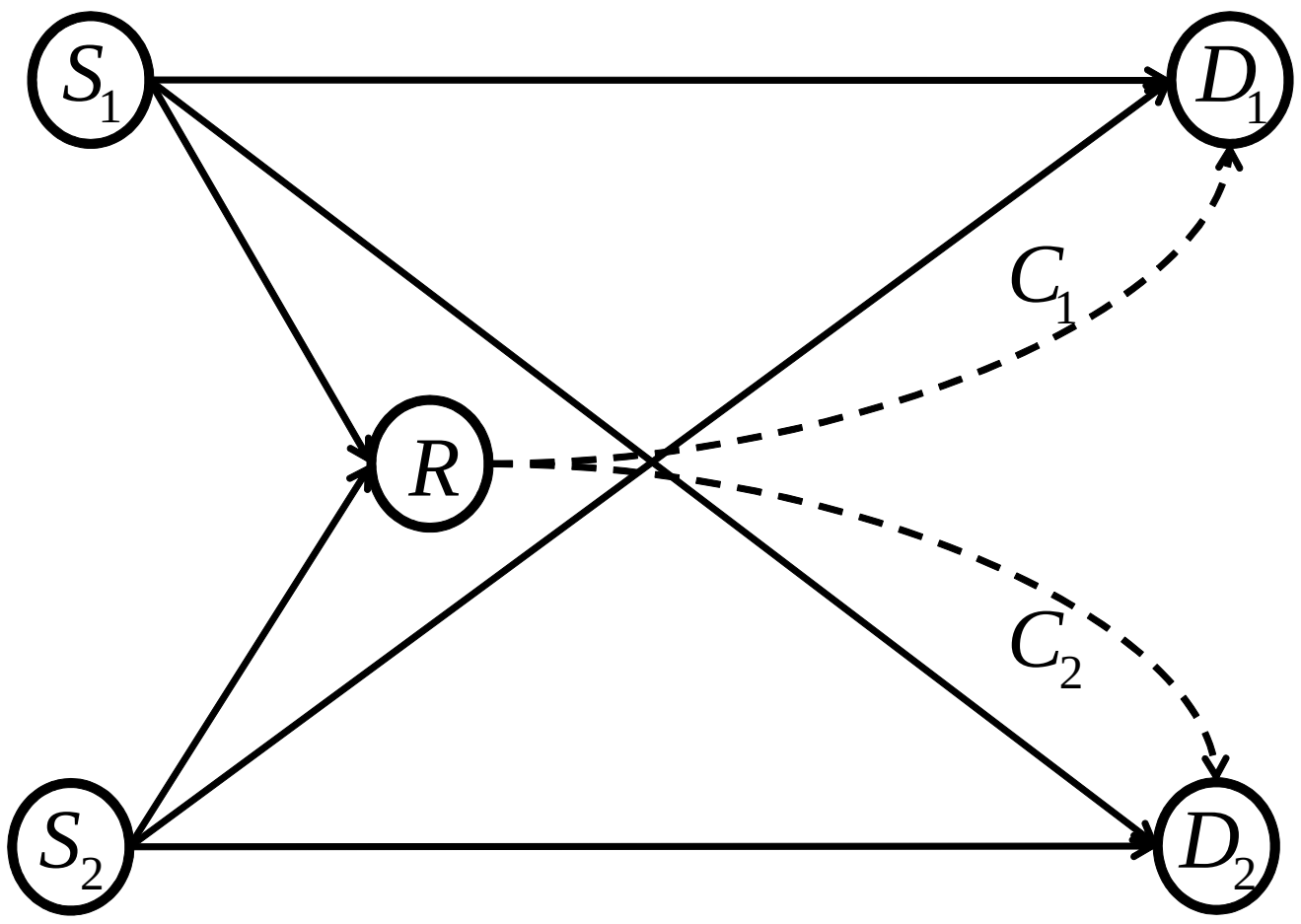

Fig. 4. Primitive interference relay channel with an oblivious relay.

*Theorem 2:* The following rate region is the capacity region for the PIFRC with an oblivious relay:

$$R_1 < I(X_1; \hat{Y}_{R1} Y_1 | X_2 Q) \tag{42}$$

$$R_1 < I(X_1; Y_1 | X_2 Q) + C_1 - I(Y_R; \hat{Y}_{R1} | X_1 X_2 Y_1 Q) \tag{43}$$

$$R_2 < I(X_2; \hat{Y}_{R2} Y_2 | X_1 Q) \tag{44}$$

$$R_2 < I(X_2; Y_2 | X_1 Q) + C_2 - I(Y_R; \hat{Y}_{R2} | X_1 X_2 Y_2 Q) \tag{45}$$

$$R_1 + R_2 < I(X_1 X_2; \hat{Y}_{R1} Y_1 | Q) \tag{46}$$

$$R_1 + R_2 < I(X_1 X_2; Y_1 | Q) + C_1 - I(Y_R; \hat{Y}_{R1} | X_1 X_2 Y_1 Q) \tag{47}$$

$$R_1 + R_2 < I(X_1 X_2; \hat{Y}_{R2} Y_2 | Q) \tag{48}$$

$$R_1 + R_2 < I(X_1 X_2; Y_2 | Q) + C_2 - I(Y_R; \hat{Y}_{R2} | X_1 X_2 Y_2 Q) \tag{49}$$

for all input distributions

$$p(q)p(x_1|q)p(x_2|q)p(\hat{y}_{R1}|y_R q)p(\hat{y}_{R2}|y_R q) \tag{50}$$

When the channel transition probability makes the following strong interference conditions hold for all input distribution $p(x_1)p(x_2)$

$$I(X_1;Y_1Y_R|X_2) \leq I(X_1;Y_2|X_2) \tag{51}$$

$$I(X_2;Y_2Y_R|X_1) \leq I(X_2;Y_1|X_1) \tag{52}$$

*Proof:* The achievability follows from the GCF relaying, which is similar to the one we used for *Theorem 1*. The only difference here is that the relay needs to generate two compression codebooks for each destination, and for completeness we provide an outline of the scheme in Appendix D.

For the outerbounds, we need to utilize the strong interference conditions (51) and (52). In fact, the conditions (51) and (52) imply that (see [20] for details)

$$I(X_1^n;Y_1^nY_R^n|X_2^nU) \leq I(X_1^n;Y_2^n|X_2^nU) \tag{53}$$

$$I(X_2^n;Y_2^nY_R^n|X_1^nU) \leq I(X_2^n;Y_1^n|X_1^nU) \tag{54}$$

Since $S_1$ and $S_2$ are functions of $Y_R^n$, we have the following lemma:

*Lemma 2:*

$$I(X_1^n;Y_1^nS_1|X_2^nU) \leq I(X_1^n;Y_1^nY_R^n|X_2^nU) \leq I(X_1^n;Y_2^n|X_2^nU) \leq I(X_1^n;Y_2^nS_2|X_2^nU) \tag{55}$$

$$I(X_2^n;Y_2^nS_2|X_1^nU) \leq I(X_2^n;Y_2^nY_R^n|X_1^nU) \leq I(X_2^n;Y_1^n|X_1^nU) \leq I(X_2^n;Y_1^nS_1|X_1^nU) \tag{56}$$

*Proof:* To prove this relation, we proceed as follows:

$$I(X_1^n;Y_1^nY_R^nS_1|X_2^nU) \tag{57}$$

$$= I(X_1^n;Y_1^nS_1|X_2^nU) + I(X_1^n;Y_R^n|X_2^nY_1^nS_1U) \tag{58}$$

$$= I(X_1^n;Y_1^nY_R^n|X_2^nU) + I(X_1^n;S_1|X_2^nY_1^nY_R^nU) \tag{59}$$

Since $(S_1, S_2) = f(Y_R^n)$, we have

$$I(X_1^n; S_1 | X_2^n Y_1^n Y_R^n U) = 0 \tag{60}$$

Consequently, we can obtain the following inequality

$$I(X_1^n; Y_1^n S_1 | X_2^n U) \leq I(X_1^n; Y_1^n Y_R^n | X_2^n U), \tag{61}$$

and the relation (55) can be obtained accordingly. The relation (56) can be obtained in the same fashion. ■

We are now ready to prove the capacity results. The outerbounds for individual rates can be obtained by setting

$$\hat{Y}_{R1,i} = S_1 X_1^{i-1} X_{1,i+1}^n X_2^{i-1} X_{2,i+1}^n Y_1^{i-1} Y_{1,i+1}^n Y_R^{i-1}, \tag{62}$$

and

$$\hat{Y}_{R2,i} = S_2 X_1^{i-1} X_{1,i+1}^n X_2^{i-1} X_{2,i+1}^n Y_2^{i-1} Y_{2,i+1}^n Y_R^{i-1}. \tag{63}$$

using similar steps as in *Theorem 1*.

For the sum rate outerbounds, we have

$$\begin{aligned}
n(R_1 + R_2) &= H(W_1) + H(W_2) &(64)\\
&= H(W_1|Q) + H(W_2|Q) &(65)\\
&\leq I(W_1; Y_1^n S_1 F_1 F_2 | W_2 Q) + I(W_2; Y_2^n S_2 F_1 F_2 | Q) + n\epsilon_n &(66)\\
&= I(W_1; F_1 F_2 | Q W_2) + I(W_1; Y_1^n S_1 | Q W_2 F_1 F_2) + I(W_2; F_1 F_2 | Q)\\
&\quad + I(W_2; Y_2^n S_2 | Q F_1 F_2) + n\epsilon_n &(67)\\
&\leq I(W_1; Y_1^n S_1 | Q F_2 W_2 F_1) + I(W_2 F_2; Y_2^n S_2 | Q F_1) + n\epsilon_n &(68)\\
&\leq I(X_1^n; Y_1^n S_1 | Q X_2^n F_1) + I(X_2^n; Y_2^n S_2 | Q F_1) + n\epsilon_n &(69)\\
&\leq I(X_1^n; Y_2^n S_2 | Q X_2^n F_1) + I(X_2^n; Y_2^n S_2 | Q F_1) + n\epsilon_n &(70)
\end{aligned}$$

$$\leq I(X_1^n X_2^n; Y_2^n S_2|Q) \tag{71}$$

where in (70) we used condition (55) with with $U = QF_1$. From this we can derive the bounds (48) and (49) following similar steps as in *Theorem 1*, and the bounds (46) and (47) can be obtained using condition (56).

With the auxiliary random variables specified by (62) and (63), the probability distribution is factorized as

$$p(q)p(x_1|q)p(x_2|q)p(\hat{y}_{R1}\hat{y}_{R2}|y_R q)p(y_1 y_2 y_R|x_1 x_2) \tag{72}$$

Note that the input distribution

$$p(q)p(x_1|q)p(x_2|q)p(\hat{y}_{R1}|y_R q)p(\hat{y}_{R2}|y_R q)p(y_1 y_2 y_R|x_1 x_2) \tag{73}$$

yields the same rate region as the one specified by (72). This is because all the rate expressions only depend on the marginal distribution

$$p(q)p(x_1|q)p(x_2|q)p(\hat{y}_{R1}|y_R q)p(y_1 y_R|x_1 x_2), \tag{74}$$

and

$$p(q)p(x_1|q)p(x_2|q)p(\hat{y}_{R2}|y_R q)p(y_2 y_R|x_1 x_2). \tag{75}$$

For each distribution factorized as (72), we can always find a distribution factorized as (73) that yields the same marginal distribution (74) and (75).

It then suffices to constrain the probability distribution to the set specified by (50). ■

*Remark 3:* The above Theorem shows that it is optimal for the relay to generate the compression codebooks for different destinations independently. The probability distribution used for generating the codebooks can be optimized independently for each destination.

*Remark 4:* Similar as PMARC, noisy network coding yields the same rate region as GCF for PIFRC. For the general IFRC, noisy network coding can have better performance, since it does not use Wyner-Ziv binning and can overcome the limitation placed on GCF by the receiver with the worst relay-destination channel. In PIFRC, however, relay has two digital links to the destinations with different capacity, and thus the rate of the bin indices in the GCF scheme is no longer limited by the receiver with the worst relay-destination channel, leading to identical rates for GCF and noisy network coding. This can be shown in a similar manner as the one for PMARC in Appendix C and is omitted for that reason.

### *B. Discussion on the Strong Interference Condition*

The strong interference condition here is different from that of the interference channel, which is of the form $I(X_1;Y_1|X_2) \leq I(X_1;Y_2|X_2)$. In the strong interference condition (55) and (56), the relay serves as a virtual antenna shared by both destinations.

To gain a more intuitive grasp of these conditions, consider the Gaussian IFRC

$$Y_1 = h_{11}X_1 + h_{21}X_2 + Z_1 \tag{76}$$

$$Y_2 = h_{12}X_1 + h_{22}X_2 + Z_2 \tag{77}$$

$$Y_R = h_{1R}X_1 + h_{2R}X_2 + Z_R, \tag{78}$$

where $Z_1, Z_2, Z_R$ are zero-mean Gaussian random variables with unit variance.

For this model, an equivalent condition for the strong interference conditions (51) and (52) is

$$h_{12}^2 \geq h_{11}^2 + h_{1R}^2 \tag{79}$$

$$h_{21}^2 \geq h_{22}^2 + h_{2R}^2, \tag{80}$$

i.e., the strength of the interference link is greater than the sum of the direct link and the corresponding source-relay link. To see this, we recognize that the left-hand-side of (51) under the Gaussian model becomes

$$I\left(X_1; \mathbf{h}X_1 + \mathbf{Z}\right) \tag{81}$$

$$= I\left(X_1; \frac{1}{\sqrt{h_{11}^2 + h_{1R}^2}}\mathbf{H}\left(\mathbf{h}X_1 + \mathbf{Z}\right)\right) \tag{82}$$

$$= I\left(X_1; \sqrt{h_{11}^2 + h_{1R}^2} X_1 + Z\right) \quad (83)$$

where

$$\mathbf{h} = \begin{pmatrix} h_{11} \\ h_{1R} \end{pmatrix} \quad (84)$$

$$\mathbf{H} = \begin{pmatrix} h_{11} & h_{1R} \\ h_{1R} & -h_{11} \end{pmatrix} \quad (85)$$

$$\mathbf{Z} = \begin{pmatrix} Z_1 \\ Z_R \end{pmatrix} \quad (86)$$

and $Z$ is a zero-mean Gaussian random variable with unit variance.

Moreover, the right-hand-side of (51) becomes

$$I(X_1; h_{12}X_1 + Z_2). \quad (87)$$

The equivalence between the conditions (51),(52) and the conditions (79),(80) are then immediate.

We can see that the strong interference condition implies that the interference link from $S_1$ to $D_2$ can provide the destination $D_2$ with more information regarding $X_1$ than the amount of information of $X_1$ contained in the combination of links $S_1$ to $D_1$ and $S_1$ to relay. Therefore the destinations are able to decode both messages.

*Remark 5:* Note that different from the interference channel, the capacity result cannot be extended to the Gaussian case directly, since the optimal input distribution is not necessarily Gaussian.

### *C. Extension to M-source K-destination Multicast Networks*

The result for the PIFRC with an oblivious relay is obtained by establishing the strong interference condition under which the channel is equivalent to a compound PMARC with an oblivious relay. This result can be extended to $M$-source $K$-destination multicast network with an oblivious relay, as defined in Section II.

*Corollary 2:* Let $\mathcal{S} \subseteq \mathcal{M}$, $d \in \mathcal{K}$, $R_{\mathcal{S}} \triangleq \sum_{i\in\mathcal{S}} R_i$. The capacity region of the $M$-source $K$-destination multicast network with an oblivious relay can be specified by the rate vector $R_{\mathcal{M}}$ satisfying

$$R_{\mathcal{S}} \leq \min_{d\in\mathcal{K}} \left\{ I(X_{\mathcal{S}}; \hat{Y}_{R,d} Y_d | X_{\mathcal{S}^C} Q) \right\} \tag{88}$$

$$R_{\mathcal{S}} \leq \min_{d\in\mathcal{K}} \left\{ I(X_{\mathcal{S}}; Y_d | X_{\mathcal{S}^C} Q) + C_d - I(\hat{Y}_{R,d}; Y_R | X_{\mathcal{M}} Y_d Q) \right\}, \tag{89}$$

for $\forall \mathcal{S}$ and all distributions

$$p(q) \left( \prod_{i=1}^{M} p(x_i|q) \right) \left( \prod_{j=1}^{K} p(\hat{y}_{R,j}|y_R q) \right) p(y_{\mathcal{K}} y_R | x_{\mathcal{M}}). \tag{90}$$

The proof of this proposition follows the same lines of reasoning as in *Theorem 1*, where the auxiliary random variable $\hat{Y}_{Ri,d}$ is defined as

$$\hat{Y}_{Ri,d} = S_d X_{\mathcal{M}}^{i-1} X_{\mathcal{M},i+1}^{n} Y_d^{i-1} Y_{d,i+1}^{n} Y_R^{i-1}. \tag{91}$$

Note that under this definition, the auxiliary random variables $\hat{Y}_{Ri,d}$ are not independent conditioned on $Y_R$ and $Q$. However, the distribution $\prod_{j=1}^{K} p(\hat{y}_{R,j}|y_R q)$ and $p(\hat{y}_{R,\mathcal{K}}|y_R q)$ yields the same rate region as discussed in the proof of *Theorem 2*. Thus it is equivalent to assume that $\hat{Y}_{Ri,d}$ are independent conditioned on $Y_R$ and $Q$.

*Remark 6:* For the multicast model, the maximum rate improvement is limited by the out-of-band link from the relay to the destinations with the smallest capacity. The average rate improvement for each source also becomes negligible as the number of sources increases, which can be resolved by introducing multiple relays, as discussed in Section III-B.

## V. Conclusion

In this paper, we have studied network information theoretic models that enjoy the cooperation of a relay node despite the relay having no information about the codebooks used by the sources. We have focused on a class of primitive multiuser networks with an oblivious relay where the relay-destination links are out-of-band with finite capacity. Different than previous work considering special cases of such networks, we have employed the generalized compress-and-forward (GCF) relaying scheme, where the compression index and source messages are decoded jointly at the destination and found it to be capacity achieving under certain conditions. Specifically, we have established the capacity region of the primitive multiple access relay channel (PMARC) with an oblivious relay by deriving new outerbounds and showing they are tight by means of GCF. We have next extended the result to establish the capacity region for the $M$-user PMARC with an oblivious relay. For the primitive interference relay channel (PIFRC) with an oblivious relay, we have established a new strong interference condition, and derived the capacity region under this condition, which again can be achieved by GCF relaying. This result is further extended to $M$-source $K$-destination multicast network with an oblivious relay.

The results obtained in this paper are intended to provide design insights towards optimal relaying strategies in wireless ad hoc networks when the codebook information is absent at the relay nodes. The capacity region of an ad hoc network with arbitrary size and topology remains open.

## Appendix A
## Proof of Lemma 1

*Proof:* We omit the subscripts $i$ in $x_i^n$ and $y_i^n$ for clarity.

$$P_{X^n|Q^n}(x^n|q^n) \tag{92}$$

$$= \sum_{f,W:\phi^n(W,f)=x^n} P(f,W|q^n) \tag{93}$$

$$= \sum_{W=1}^{2^{nR}} \sum_{f:\phi^n(W,f)=x^n} P(f|q^n)P(W) \tag{94}$$

$$= \sum_{f:\phi^n(1,f)=x^n} P(f|q^n) \tag{95}$$

$$= \prod_{i=1}^{n} P(x_i|q_i) \sum_{x^n(2)\cdots x^n(2^{nR})} \prod_{i=1}^{n} \prod_{l=2}^{2^{nR}} P(x_i(l)|q_i) \tag{96}$$

$$= \prod_{i=1}^{n} P(x_i|q_i) \sum_{x_2^n(2)\cdots x_2^n(2^{nR})} \sum_{x_1(2)\cdots x_1(2^{nR})} \left( \prod_{l=2}^{2^{nR}} P(x_1(l)|q_1) \prod_{i=2}^{n} \prod_{l=2}^{2^{nR}} P(x_i(l)|q_i) \right) \tag{97}$$

$$= \prod_{i=1}^{n} P(x_i|q_i) \sum_{x_2^n(2)\cdots x_2^n(2^{nR})} \left( \sum_{x_1(2)\cdots x_1(2^{nR})} \prod_{l=2}^{2^{nR}} P(x_1(l)|q_1) \right) \prod_{i=2}^{n} \prod_{l=2}^{2^{nR}} P(x_i(l)|q_i) \tag{98}$$

For the middle term in the above expression, we have

$$\sum_{x_1(2)\cdots x_1(2^{nR})} \prod_{l=2}^{2^{nR}} P(x_1(l)|q_1) \tag{99}$$

$$= \sum_{x_1(3)\cdots x_1(2^{nR})} \sum_{x_1(2)} P(x_1(2)|q_1) \prod_{l=3}^{2^{nR}} P(x_1(l)|q_1) \tag{100}$$

$$= \sum_{x_1(3)\cdots x_1(2^{nR})} \left( \sum_{x_1(2)} P(x_1(2)|q_1) \right) \prod_{l=3}^{2^{nR}} P(x_1(l)|q_1). \tag{101}$$

Now it is easy to see that $\sum_{x_1(2)} P(x_1(2)|q_1) = 1$. Continue with this approach, we have

$$\sum_{x_1(2)\cdots x_1(2^{nR})} \prod_{l=2}^{2^{nR}} P(x_1(l)|q_1) = 1. \tag{102}$$

Substitute (102) into (98), and proceed with the rest of the terms, we have

$$P_{X^n|Q^n}(x^n|q^n) = \prod_{i=1}^{n} P_{X|Q}(x_i|q_i) \tag{103}$$

Using this result, we have

$$P_{Y^n|Q^n}(y^n|q^n) \tag{104}$$

$$= \sum_{x^n} P_{X^nY^n|Q^n}(x^n y^n|q^n) \tag{105}$$

$$= \sum_{x^n} P_{X^n|Q^n}(x^n|q^n) \prod_{i=1}^{n} P_{Y|X}(y_i|x_i) \tag{106}$$

$$= \sum_{x^n} \prod_{i=1}^{n} P_{X|Q}(x_i|q_i) P_{Y|X}(y_i|x_i) \tag{107}$$

$$= \sum_{x^n} \prod_{i=1}^{n} P_{XY|Q}(x_i y_i|q_i) \tag{108}$$

$$= \prod_{i=1}^{n} \sum_{x_i} P_{XY|Q}(x_i y_i | q_i) \tag{109}$$

$$= \prod_{i=1}^{n} P_{Y|Q}(y_i | q_i) \tag{110}$$

■

## APPENDIX B

## PROOF OF THEOREM 1

*Proof:* **Achievability**: The achievability of the rate region in *Theorem 1* can be obtained using GCF. The main idea is similar to CF scheme originated from [2]. The difference lies in how decoding is performed, where in GCF, the decoders jointly decode the source messages and the bin index. The detailed scheme is described in the sequel.

*Codebook generation:* Fix a distribution $p(q)\,p(x_1|q)\,p(x_2|q)\,p(\hat{y}_R|y_R q)$. First generate a sequence $q^n$ according to $\prod_{j=1}^{n} p_Q(q_j)$, which is revealed to all nodes. Based on $q^n$, for each message $w_i \in \{1, \cdots, 2^{nR_i}\}$, generate codeword $x_i^n(w_i)$ according to $\prod_{j=1}^{n} p_{X_i|Q}(x_{ij}|q_j)$ $(i = 1, 2)$. Based on $q^n$, for each $l \in \{1, \cdots, 2^{n\hat{R}}\}$, generate $\hat{y}_R^n(l)$ according to $\prod_{j=1}^{n} p_{\hat{Y}_R|Q}(\hat{y}_{Rj}|q_j)$. Partition the set $\{1, \cdots, 2^{n\hat{R}}\}$ into $2^{nC}$ bins of equal size, and denote the $k$th bin as $\mathcal{B}_k$.

*Encoding:* The sources and the relay share the time-sharing sequence $q^n$. Source 1 encodes $w_1$ into $x_1^n(w_1)$, sources 2 encodes $w_2$ into $x_2^n(w_2)$, and then the sources send the codewords into the channel. The relay receives $y_R^n$, and it looks for an index $l \in \{1, \cdots, 2^{n\hat{R}}\}$ such that $(y_R^n, \hat{y}_R^n(l), q^n) \in T_\epsilon^n$. If no such $l$ exists, the relay declares an error. If there is more than one such $l$, the relay chooses the smallest one. Assume $l \in \mathcal{B}_k$. The relay sends the bin index $k$ containing $l$ to the destination using the out-of-band link.

*Decoding:* Since the relay-destination link has capacity $C$, the bin index $k$ can be decoded correctly at the destination. Destination receives $y^n$ at the end of transmission. The decoder tries to find $\hat{w}_1$ and $\hat{w}_2$ such that $\left(q^n, x_1^n(\hat{w}_1), x_2^n(\hat{w}_2), \hat{y}_R^n(\hat{l}), y^n\right) \in T_\epsilon^n$ for some $\hat{l} \in \mathcal{B}_k$.

*Error Analysis:* Assume $w_1 = 1, w_2 = 1, l = L$. Define the following error events.

$\mathcal{E}_0 := \left\{\left(Q^n, Y_R^n, \hat{Y}_R^n(l)\right) \notin T_\epsilon^n, \forall l\right\}$

$$\mathcal{E}_1 := \left\{ \left( Q^n, X_1^n\left(1\right), X_2^n\left(1\right), \hat{Y}_R^n\left(L\right), Y^n \right) \notin \mathcal{T}_\epsilon^n \right\}$$
$$\mathcal{E}_2 := \left\{ \exists \hat{w}_1 \neq 1 : \left( Q^n, X_1^n\left(\hat{w}_1\right), X_2^n\left(1\right), \hat{Y}_R^n\left(L\right), Y^n \right) \in \mathcal{T}_\epsilon^n \right\}$$
$$\mathcal{E}_3 := \left\{ \exists \hat{w}_2 \neq 1 : \left( Q^n, X_1^n\left(1\right), X_2^n\left(\hat{w}_2\right), \hat{Y}_R^n\left(L\right), Y^n \right) \in \mathcal{T}_\epsilon^n \right\}$$
$$\mathcal{E}_4 := \left\{ \exists \hat{w}_1 \neq 1, \hat{w}_2 \neq 1 : \left( Q^n, X_1^n\left(\hat{w}_1\right), X_2^n\left(\hat{w}_2\right), \hat{Y}_R^n\left(L\right), Y^n \right) \in \mathcal{T}_\epsilon^n \right\}$$
$$\mathcal{E}_5 := \left\{ \exists \hat{w}_1 \neq 1 : \left( Q^n, X_1^n\left(\hat{w}_1\right), X_2^n\left(1\right), \hat{Y}_R^n(\hat{l}), Y^n \right) \in \mathcal{T}_\epsilon^n, \text{for some } \hat{l} \in \mathcal{B}_k, \hat{l} \neq L \right\}$$
$$\mathcal{E}_6 := \left\{ \exists \hat{w}_2 \neq 1 : \left( Q^n, X_1^n\left(1\right), X_2^n\left(\hat{w}_2\right), \hat{Y}_R^n(\hat{l}), Y^n \right) \in \mathcal{T}_\epsilon^n, \text{for some } \hat{l} \in \mathcal{B}_k, \hat{l} \neq L \right\}$$
$$\mathcal{E}_7 := \Big\{ \exists \hat{w}_1 \neq 1, \hat{w}_2 \neq 1 : \left( Q^n, X_1^n\left(\hat{w}_1\right), X_2^n\left(\hat{w}_2\right), \hat{Y}_R^n(\hat{l}), Y^n \right) \in \mathcal{T}_\epsilon^n, \text{for some } \hat{l} \in \mathcal{B}_k, \hat{l} \neq L \Big\}$$

The error probability is

$$P\left(\mathcal{E}\right) = P\left( \mathcal{E}_0 \bigcup \cup_{i=1}^7 \mathcal{E}_i \right) \tag{111}$$
$$\leq P\left(\mathcal{E}_0\right) + P\left(\mathcal{E}_1\right) + \sum_{i=2}^{7} P\left( \mathcal{E}_i \bigcap \left( \mathcal{E}_0^c \cap \mathcal{E}_1^c \right) \right) \tag{112}$$

According to covering lemma and conditional typicality lemma [17], $P\left(\mathcal{E}_0\right) \to 0$ as $n \to \infty$ as long as

$$\hat{R} > I\left( Y_R; \hat{Y}_R | Q \right) \tag{113}$$

According to conditional typicality lemma [17], $P\left(\mathcal{E}_1\right) \to 0$ as $n \to \infty$. Following the derivation in [17], we can show that $P\left(\mathcal{E}_2\right)$, $P\left(\mathcal{E}_3\right)$, $P\left(\mathcal{E}_4\right)$, $P\left(\mathcal{E}_5\right)$, $P\left(\mathcal{E}_6\right)$, $P\left(\mathcal{E}_7\right) \to 0$ as $n \to \infty$ as long as

$$R_1 < I(X_1; \hat{Y}_R Y | X_2 Q) \tag{114}$$
$$R_2 < I(X_2; \hat{Y}_R Y | X_1 Q) \tag{115}$$
$$R_1 + R_2 < I(X_1 X_2; \hat{Y}_R Y | X_2 Q) \tag{116}$$
$$R_1 + \hat{R} - C < I(X_1; Y | X_2 Q) + I(\hat{Y}_R; X_1 X_2 Y | Q) \tag{117}$$
$$R_2 + \hat{R} - C < I(X_2; Y | X_1 Q) + I(\hat{Y}_R; X_1 X_2 Y | Q) \tag{118}$$
$$R_1 + R_2 + \hat{R} - C < I(X_1 X_2; Y | Q) + I(\hat{Y}_R; X_1 X_2 Y | Q) \tag{119}$$

Combining these constraints with (113) yields the desired result.

**Sum rate upperbound**: For the sum rate, we have

$$\begin{aligned}
n(R_1+R_2) &= H(W_1W_2) && (120)\\
&= H(W_1W_2|Q') && (121)\\
&\le I(W_1W_2;Y^nSF_1F_2|Q') + n\epsilon_n && (122)\\
&= I(W_1W_2;F_1F_2|Q'W_2) + I(W_1W_2;Y^nS|Q'F_1F_2) + n\epsilon_n && (123)\\
&\le I(F_1W_1F_2W_2;Y^nS|Q') + n\epsilon_n && (124)\\
&\le I(X_1^nX_2^n;Y^nS|Q') + n\epsilon_n && (125)
\end{aligned}$$

We can further bound (125) in two different ways.

$$\begin{aligned}
&I(X_1^nX_2^n;Y^nS|Q') && (126)\\
&= H(X_1^nX_2^n|Q') - H(X_1^nX_2^n|Y^nSQ') && (127)\\
&= \sum_{i=1}^{n} H(X_{1i}X_{2i}|Q') - \sum_{i=1}^{n} H(X_{1i}X_{2i}|Y^nSQ'X_1^{i-1}X_2^{i-1}) && (128)\\
&\le \sum_{i=1}^{n} H(X_{1i}X_{2i}|Q') - \sum_{i=1}^{n} H(X_{1i}X_{2i}|SX_1^{i-1}X_{1,i+1}^{n}X_2^{i-1}X_{2,i+1}^{n}Y^{i-1}Y_{i+1}^{n}Y_R^{i-1}Y_iQ') && (129)\\
&= \sum_{i=1}^{n} H(X_{1i}X_{2i}|Q') - \sum_{i=1}^{n} H(X_{1i}X_{2i}|\hat{Y}_{Ri}Y_iQ') && (130)\\
&= \sum_{i=1}^{n} I(X_{1i}X_{2i};\hat{Y}_{Ri}Y_i|Q') && (131)
\end{aligned}$$

Or

$$\begin{aligned}
&I(X_1^nX_2^n;Y^nS|Q') && (132)\\
&= I(X_1^nX_2^n;Y^n|Q') + I(X_1^nX_2^n;S|Q'Y^n) && (133)\\
&= H(Y^n|Q') - H(Y^n|Q'X_2^nX_1^n) + H(S|Q'Y^n) - H(S|Q'X_1^nX_2^nY^n) && (134)\\
&\le \sum_{i=1}^{n} H(Y_i|Q') - \sum_{i=1}^{n} H(Y_i|X_{1i}X_{2i}Q') + H(S)
\end{aligned}$$

$$- \left(H(S|Q'X_1^nX_2^nY^n) - H(S|Q'X_1^nX_2^nY^nY_R^n)\right) \tag{135}$$

$$\leq \sum_{i=1}^{n} I(X_{1i}X_{2i};Y_i|Q') + nC - I(S;Y_R^n|Q'X_1^nX_2^nY^n) \tag{136}$$

$$= \sum_{i=1}^{n} I(X_{1i}X_{2i};Y_i|Q') + nC - \sum_{i=1}^{n} I(S;Y_{Ri}|Q'X_1^nX_2^nY^nY_R^{i-1}) \tag{137}$$

$$= \sum_{i=1}^{n} I(X_{1i}X_{2i};Y_i|Q') + nC - \sum_{i=1}^{n}\Big(H(Y_{Ri}|Q'X_1^nX_2^nY^nY_R^{i-1}) - H(Y_{Ri}|SQ'X_1^nX_2^nY^nY_R^{i-1})\Big) \tag{138}$$

$$= \sum_{i=1}^{n} I(X_{1i}X_{2i};Y_i|Q') + nC - \sum_{i=1}^{n}\Big(H(Y_{Ri}|Q'X_{1i}X_{2i}Y_i) - H(Y_{Ri}|SX_1^{i-1}X_{1,i+1}^nX_2^{i-1}X_{2,i+1}^nY^{i-1}Y_{i+1}^nY_R^{i-1}X_{1i}X_{2i}Y_iQ')\Big) \tag{139}$$

$$= \sum_{i=1}^{n} I(X_{1i}X_{2i};Y_i|Q') + nC - \sum_{i=1}^{n} I(\hat{Y}_{Ri};Y_{Ri}|X_{1i}X_{2i}Y_iQ') \tag{140}$$

The result can be obtained by introducing another time sharing random variable $Q'' \sim \mathcal{U}(\{1,2,\cdots,n\})$ and setting $Q = (Q'', Q')$. The way we define the random variable $\hat{Y}_{Ri}$ implies the distribution (16). ■

## Appendix C

## Comparison between GCF and Noisy Network Coding

For the general multiple access relay channel (MARC), the achievable rate region provided by noisy network coding is given by [19]

$$R_1 < I(X_1;\hat{Y}_RY|X_2X_R) \tag{141}$$

$$R_1 < I(X_1X_R;Y|X_2) - I(Y_R;\hat{Y}_R|X_1X_2X_RY) \tag{142}$$

$$R_2 < I(X_2;\hat{Y}_RY|X_1X_R) \tag{143}$$

$$R_2 < I(X_2X_R;Y|X_1) - I(Y_R;\hat{Y}_R|X_1X_2X_RY) \tag{144}$$

$$R_1 + R_2 < I(X_1X_2;\hat{Y}_RY|X_R) \tag{145}$$

$$R_1 + R_2 < I(X_1X_2X_R;Y) - I(Y_R;\hat{Y}_R|X_1X_2X_RY) \tag{146}$$

for all distributions

$$p(x_1)p(x_2)p(x_R)p(\hat{y}_R|y_R x_R)p(y y_R|x_1 x_2 x_R) \tag{147}$$

For the PMARC, we first set $Y = (Y', Y'')$. The probability distribution (147) reduces to the following form:

$$p(x_1)p(x_2)p(\hat{y}_R|y_R x_R)p(y' y_R|x_1 x_2)p(x_R)f(y''|x_R) \tag{148}$$

where $f(y''|x_R)$ is a deterministic function, and $y''$ is the signal received at the destination from the digital relay-destination link. The digital link has capacity $C$, i.e., $H(Y'') = C$.

The rate expression (141) becomes

$$R_1 < I(X_1; \hat{Y}_R Y' Y''|X_2 X_R) \tag{149}$$

$$= I(X_1; \hat{Y}_R Y'|X_2 X_R) \tag{150}$$

The rate expression (142) can be written as follows:

$$I(X_1 X_R; Y' Y''|X_2) - I(Y_R; \hat{Y}_R|X_1 X_2 X_R Y' Y'') \tag{151}$$

$$= I(X_1 X_R; Y|X_2) + I(X_1 X_R; Y''|X_2 Y') - I(Y_R; \hat{Y}_R|X_1 X_2 X_R Y') \tag{152}$$

$$= I(X_1; Y'|X_2) + C - I(Y_R; \hat{Y}_R|X_1 X_2 X_R Y') \tag{153}$$

The derivations follow from the probability distribution (148). We can obtain the same rate expressions as in (10) and (11) by setting $Y' = Y$ and removing $X_R$ due to the fact that the link is digital. Using similar method for the rate expressions (143)$-$(146), we can obtain exactly the same rate region as the one obtained by using GCF scheme in *Theorem 1*.

## APPENDIX D

## PROOF OF THEOREM 2

*Proof:* **Achievability**: To show the achievability of the rate region in *Theorem 2*, we use GCF scheme at the relay to convey different compression indices to each destination. The encoding and decoding process is similar to that from *Theorem 1*. The only difference is how the compression at the relay is performed. Specifically, we fix a distribution

$$p(q)p(x_1|q)p(x_2|q)p(\hat{y}_{R1}|y_R q)p(\hat{y}_{R2}|y_R q), \tag{154}$$

and generate two compression codebooks $\hat{Y}_{Ri}^n(l_i)$ with $l_i \in \{1, 2, \cdots, 2^{n\hat{R}_i}\}, i = 1, 2$ at the relay based on the distributions $\prod_{j=1}^n p_{\hat{Y}_{Ri}|Q}(\hat{y}_{Ri,j}|q_j)$. We then partition the set $\{1, \cdots, 2^{n\hat{R}_i}\}$ into $2^{nC_i}$ bins of equal size, and denote the $k$th bin of the $i$th codebook as $\mathcal{B}_{ki}$. Based on the received sequence, the relay chooses two compression indices from each compression codebook such that $(\hat{y}_{R1}^n(l_1), y_R^n, q^n) \in T_\epsilon^n\left(\hat{Y}_{R1}, Y_R|Q^n\right)$ and $(\hat{y}_{R2}^n(l_2), y_R^n, q^n) \in T_\epsilon^n\left(\hat{Y}_{R2}, Y_R|Q^n\right)$. We then find the bin indices containing the compression indices, i.e., $l_1 \in \mathcal{B}_{k_1}$ and $l_2 \in \mathcal{B}_{k_2}$. The bin index $k_1$ is relayed to destination 1, and the bin index $k_2$ is relayed to destination 2. The destinations jointly decode the intended compression indices and the source messages. Note that in the decoding process, as long as the decoding of the intended source message is correct, there is no decoding error regardless the decoding of the non-intended source message is correct or not. The rate region in *Theorem 2* is then the intersection of the rate regions for PMARC with an oblivious relay at each destination, where the rate constraints for the non-intended source messages are removed. ■